\documentclass[10pt, secnumarabic, amssymb, amsmath, aps, pra, twocolumn, superscriptaddress]{revtex4-2}
\usepackage{graphicx}
\usepackage{amsmath}
\usepackage{physics}
\usepackage{color}

\usepackage{natbib}
\usepackage{ulem}
\graphicspath{{Figures/}}

\newcommand{\FX}{\ensuremath{\mathrm{FX}}}
\newcommand{\DO}{\ensuremath{\mathrm{D^{0}}}}
\newcommand{\DOX}{\ensuremath{\mathrm{D^{0}X}}}
\newcommand{\XX}{\ensuremath{\mathrm{X-}}}

\begin{document}

\title{Single quantum emitters with spin ground states based on Cl bound excitons in ZnSe}

\author{Aziz Karasahin}
\email{azizk@umd.edu}
\affiliation{Institute for Research in Electronics and Applied Physics, Department of Electrical and Computer Engineering and Joint Quantum Institute, University of Maryland, College Park, MD 20742}

\author{Robert M. Pettit}
\affiliation{Institute for Research in Electronics and Applied Physics, Department of Electrical and Computer Engineering and Joint Quantum Institute, University of Maryland, College Park, MD 20742}
\affiliation{Intelligence Community Postdoctoral Research Fellowship Program, University of Maryland, College Park, MD 20742}

\author{Nils von den Driesch}
\affiliation{Peter-Gr\"{u}nberg-Institute (PGI-9), Forschungszentrum J\"{u}lich GmbH, 52425 J\"{u}lich, Germany \& JARA-FIT (Fundamentals of Future Information Technology, J\"{u}lich-Aachen Research Alliance, 52062 Aachen, Germany}

\author{Marvin Marco Jansen}
\affiliation{Peter-Gr\"{u}nberg-Institute (PGI-9), Forschungszentrum J\"{u}lich GmbH, 52425 J\"{u}lich, Germany \& JARA-FIT (Fundamentals of Future Information Technology, J\"{u}lich-Aachen Research Alliance, 52062 Aachen, Germany}

\author{Alexander Pawlis}
\affiliation{Peter-Gr\"{u}nberg-Institute (PGI-9), Forschungszentrum J\"{u}lich GmbH, 52425 J\"{u}lich, Germany \& JARA-FIT (Fundamentals of Future Information Technology, J\"{u}lich-Aachen Research Alliance, 52062 Aachen, Germany}

\author{Edo Waks}
\email{edowaks@umd.edu}
\affiliation{Institute for Research in Electronics and Applied Physics, Department of Electrical and Computer Engineering and Joint Quantum Institute, University of Maryland, College Park, MD 20742}


\begin{abstract}
Defects in wide-bandgap semiconductors are promising qubit candidates for quantum communication and computation. Epitaxially grown II-VI semiconductors are particularly promising host materials due to their direct bandgap and potential for isotopic purification to a spin-zero nuclear background. Here, we show a new type of single photon emitter with potential electron spin qubits based on Cl impurities in ZnSe. We utilize a quantum well to increase the binding energies of donor emission and confirm single photon emission with short radiative lifetimes of 192 ps. Furthermore, we verify that the ground state of the Cl donor complex contains a single electron by observing two-electron satellite emission, leaving the electron in higher orbital states. We also characterize the Zeeman splitting of the exciton transition by performing polarization-resolved magnetic spectroscopy on single emitters. Our results suggest single Cl impurities are suitable as single photon source with potential photonic interface.
\end{abstract}

\date{Compiled \today}

\maketitle





\section{INTRODUCTION}

Impurities \cite{Awschalom2018} in wide-bandgap semiconductors are an essential building block for quantum photonic devices \cite{Wolfowicz2021,Awschalom2018,Weber2010}. These isolated impurities generate single photons that exhibit transform-limited linewidths \cite{Trusheim2020,Bradac2019} and small inhomogeneous broadening \cite{Sanaka2009,Sanaka2012} through their radiative optical transitions. They also provide isolated bound electrons \cite{Linpeng2018} or holes \cite{Strauf2002} that can serve as spin-photon interfaces with long coherence times \cite{Sukachev2017,Linpeng2021}. Furthermore, semiconductor fabrication and integration methods allow efficient spin-photon interfaces \cite{Atature2018,Luo2019,Javadi2018}. Hence, these platforms could satisfy major requirements of scalable quantum technology by combining pristine radiative properties with long-coherence time spin qubits in a practical setting. 

Zinc Selenide (ZnSe) is appealing material as the host crystal because it possesses a direct optical bandgap and large natural abundances of nuclear spin-0 isotopes. The direct band gap allows efficient radiative transitions that can generate bright excitonic emission. Both single fluorine (F) donors \cite{DeGreve2010} and single nitrogen  (N) acceptors \cite{Strauf2002} have been optically isolated in this host material. Ensemble measurements have also revealed the potential for optical active spin qubits \cite{Greilich2012}. Recently, the growth of isotopically purified (Zn,Mg)Se/ZnSe quantum well structures was also demonstrated \cite{Pawlis2019}. Spin purification of the host crystal further resulted in extended spin coherence times \cite{Kirstein2021}. Moreover, donor-bound excitons in nanostructured pillars have been used to demonstrate indistinguishable single photon emission between two independent emitters \cite{Sanaka2009}, two-photon entanglement \cite{Sanaka2012}, and optical pumping of the donor spin \cite{Sleiter2013}.

Most studies on impurity-bound exciton emission have focused on the fluorine (F) impurity measurements, which possesses a smaller atomic radius compared to selenium (Se) atoms. The small radius causes self-compensation and limits the doping concentrations that can be achieved. On the other hand, chlorine (Cl) has an atomic radius closer to that of Se and is therefore more suitable as dopant with a low probability of self-compensation \cite{Ohkawa1987}. This enables Cl doping levels that cover a wide range of densities between $10^{16}$~cm$^{-3}$ and $10^{19}$~cm$^{-3}$. But all past work on Cl impurity bound exciton emission was in the large ensemble regime, and single Cl donor bound exciton emission has yet to be studied.

\section{RESULTS}
\begin{figure*}
	\centering
	\includegraphics{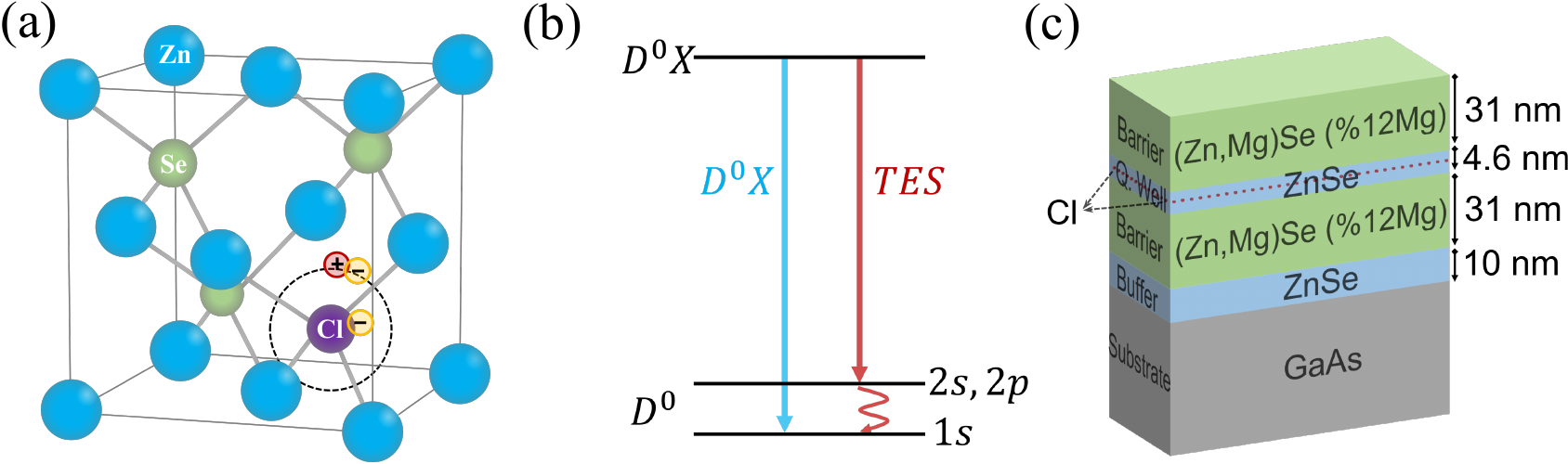}
	\caption{(a) Crystal structure of ZnSe with a single Cl impurity replacing a Se atom (b) Energy band diagram showing the ground state (\DO) and excited state (\DOX), along with the bound exciton and two-electron-satellite emission, shown as blue and red arrow (c) Layer structure of the grown material with in-situ doped Cl impurities in the ZnSe quantum well.}
	\label{fig1}
\end{figure*}
Here we isolate single Cl impurities in ZnSe and show that they act as an efficient single photon source with a potential to act as a spin qubit. Optical isolation of single bound excitons requires a sufficiently low density of impurity atoms and sufficiently large exciton binding energy to separate bound exciton emission from the free exciton emission. We achieve these requirements by using a quantum well that is delta-doped with a low concentration of Cl impurities. Quantum confinement in the order of Bohr radius of bound exciton results in increase in the binding energy of bound excitons, hence, allows us to better separate the free exciton and bound exciton emission in energy \cite{Mariette1988,Kavokin1992,Pawlis2011}. We show fast and stable single photon emission from single Cl donor bound excitons. We also confirm the existence of a single electron ground state by observing two-electron satellite transitions and observing the Zeeman splitting by Faraday magnetic field. These results establish Cl impurities in ZnSe as a promising optically active qubit system.

Figure \ref{fig1}.a illustrates the atomic structure of a ZnSe crystal with a single Cl impurity. In its intrinsic form, zinc and selenium atoms are arranged in a face centered cubic lattice where each zinc atom is surrounded by 4 selenium atoms. A single Cl impurity replaces a Se atom \cite{Poykko1998} when chlorine impurities are introduced into the lattice via in-situ doping or ion implantation. The Cl impurity serves as a neutral electron donor, consisting of the positively charged chlorine core and an additional single bound electron, shown in Figure \ref{fig1}.a This extra donor-bound electron may serve as a stable single photon emitter with an electron spin ground state. 

Figure \ref{fig1}.b shows the level structure of Cl impurity. Optical excitation above the ZnSe band gap produces free excitons in the quantum well. These excitons can become bound to the Cl impurity, forming an impurity bound exciton state \DOX. This state can radiatively combine back to the $1s$ ground state, emitting a photon around 440 nm. The bound exciton can also recombine to the higher energy $2s,2p$ states through an Auger recombination process, emitting a longer wavelength photon. The observation of two-electron-satellite establishes the presence of an electron ground state.

Figure \ref{fig1}.c shows the epitaxial layer structure of the sample used in this work. The sample is composed of a 4.6 nm ZnSe quantum well embedded in a (Zn$_{0.88}$Mg$_{0.12}$)Se barrier on top and bottom each with a thickness of 31 nm. To achieve controlled doping of impurities while sustaining the high crystalline quality, we grow a thin Cl delta-doped layer in the middle of the ZnSe quantum well. The delta-doped layer has a Cl sheet concentration of approximately $10^{9}$~cm$^{-2}$. The overall structure was grown on a GaAs substrate covered with a 10 nm ZnSe buffer layer.

We performed optical measurements with a home-built confocal microscope. We mounted samples on an XYZ piezo stage and maintained the sample temperature at 3.6K in a closed-loop helium cryostat. A 405 nm wavelength continuous wave diode laser, whose energy is larger than the bandgaps of the barrier and the quantum well, was used for excitation. We analyzed the photoluminescence signal with an imaging spectrometer (Princeton Instruments, SP-2750) with a spectral resolution of 15 pm or single photon detectors (Micro Photon Devices, PDM series). To optically isolate individual impurities in the bulk material, we used confocal collection to a single-mode fiber.

\begin{figure*}
	\centering
	\includegraphics{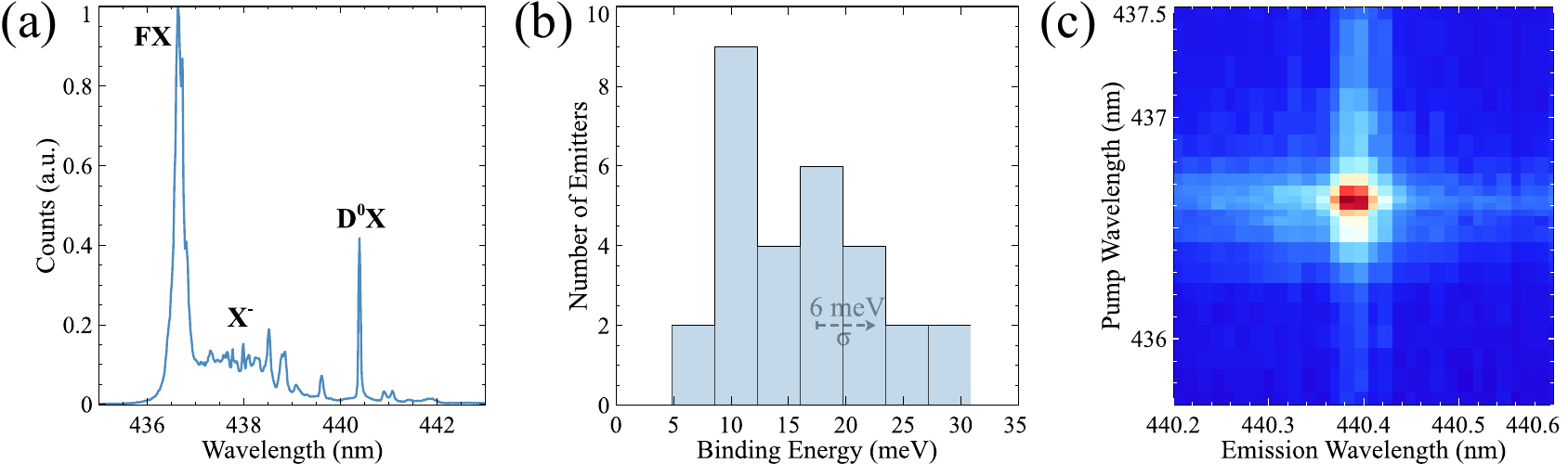}
	\caption{(a) PL spectrum obtained with above band excitation, demonstrating free exciton (\FX), trion (\XX) and bound exciton (\DOX) emission (b) Histogram showing the distribution of binding energies of \DOX~lines, with an average binding energy of 15 meV and a standard deviation of 6 meV(c) Photoluminescence excitation spectrum of bound exciton line.}
	\label{fig2}
\end{figure*}

\begin{figure}
	\centering
	\includegraphics[width=\columnwidth]{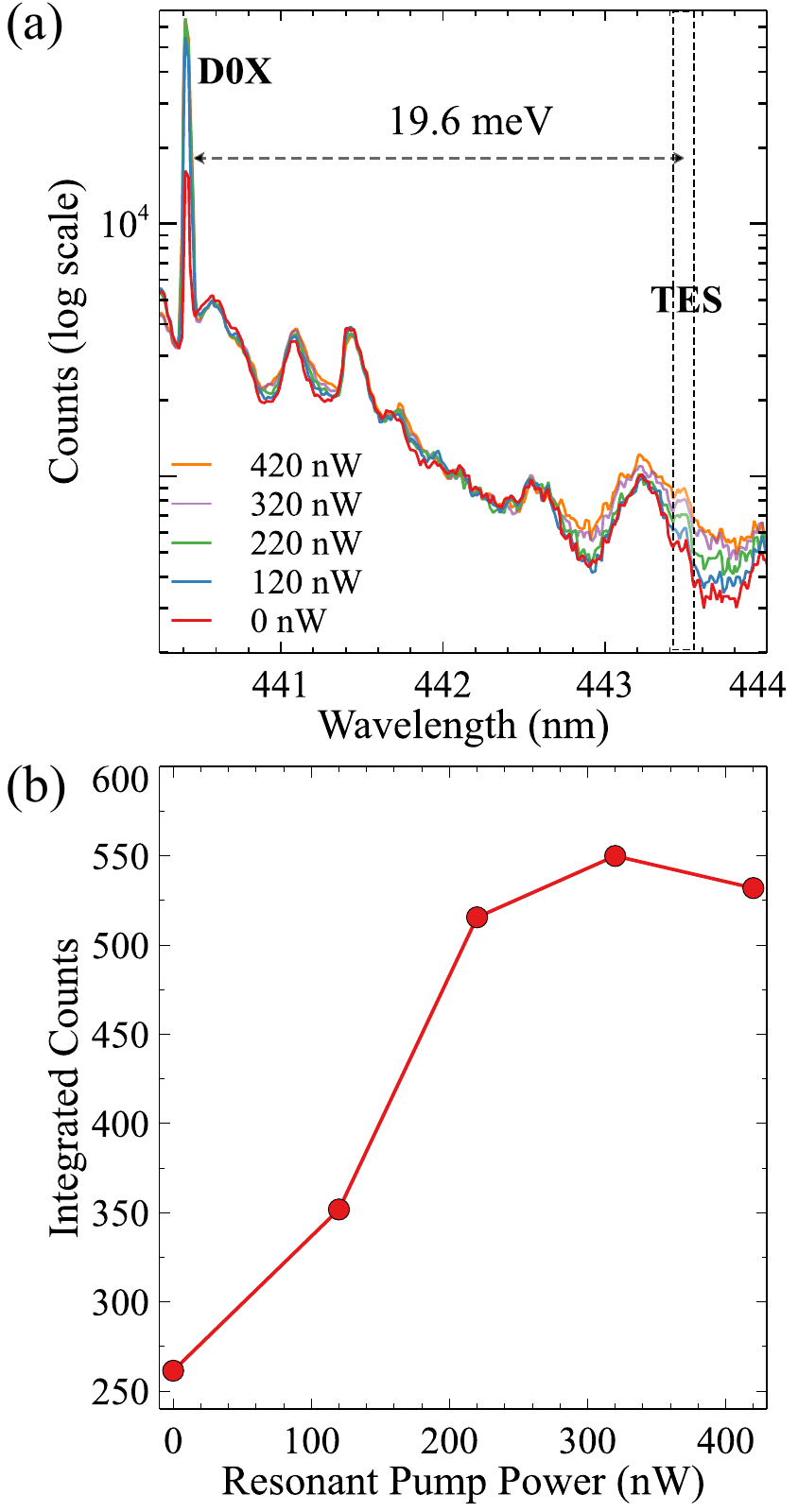}
	\caption{(a) Spectrums showing the TES line with increasing resonant pump power, (b) Background corrected and integrated counts from dashed spectral window showing saturation of TES line with increasing pump power.}
	\label{fig3}
\end{figure}

The photoluminescence spectrum from the sample is shown in Figure \ref{fig2}.a We observed broad emission from the heavy-hole free exciton (labeled as \FX, centered at 436.7 nm) and negatively charged trion (\XX) in the quantum well \cite{Homburg2000}. In addition to these lines, we observed a sharp localized emission line at 440.4 nm from excitons bound to neutral donors (\DOX). As we move the position of the sample, we observe different localized lines appear in the spectrum at different spectral locations. In contrast, the energy of free exciton and trion lines remain constant throughout the sample.

Figure \ref{fig2}.b plots a histogram of the energies of 29 emission lines acquired at different locations in the sample relative to the free exciton emission. The mean distribution of the emissions is shifted by 15 meV relative to the free exciton line. This binding energy is reasonable assuming similar Bohr radius (about 3 nm) of free exciton and bound exciton, leading to about 2-3 times increased \DOX~binding energy compared to reported bulk values of 7 meV \cite{Dean1981}. Using the distribution of emission energies, we calculate the standard deviation of the binding energy as 6 meV. The observed inhomogeneous broadening is possibly caused by fluctuations of the donor positions within the well, local thickness fluctuations of the quantum well and local strain variations \cite{Osario1988,Senger2004}.

To verify that the observed \DOX~emission originates from a bound state of the quantum well free exciton, we perform a photoluminescence excitation measurement. We use a tunable laser with a fixed pump power of 100 nW.  We tune the laser emission over the free exciton line and monitor the emission from a \DOX line. In the absence of above band laser, moderately n-doped quantum well has been shown to exhibit low-density hole gas. Applying an appropriate above-barrier laser supplies the sufficient electron concentration in the quantum well \cite{Puls2001,Hayne1994,Zhukov2014}. During this measurement, we inject a weak above-band laser (about 3 nW at 405 nm) in addition to pump laser. At this power, the above-band laser produces negligible photoluminescence, however, it helps adjusting the carrier densities in the quantum well.  Figure \ref{fig2}.c shows the photoluminescence excitation spectrum of a \DOX~line as we tune the pump laser over the free exciton line. We observe strong emission from the \DOX~ line when the laser is resonant with the free exciton. This enhancement suggests that free excitons in the quantum well non-radiatively relax to form donor-bound excitons at the impurity.

Figure \ref{fig3} shows the observation of two-electron satellite emission in the presence of resonant excitation. The top panel shows the PL signal obtained with a weak above-band laser and varying power of resonant pump laser tuned at the \DOX~ line. Two-electron-satellite (TES) emission 19.6 meV below the \DOX~ line emerges, and its intensity increases as we increase the resonant pump power, shown in Figure \ref{fig3}.a Satellite emission appears lower in energy by an amount equal to the spacing of the n=1 and n=2 of \DO~ electron orbitals determined by a Hydrogenic donor model \cite{Dean1981}. The weak nature of this emission is due to the predominant relaxation of \DOX~ state to 1s ground state. However, there is a finite probability of relaxation to any of the excited (2s, 2p, etc.) states as discussed in the Figure \ref{fig1}.b To be able to observe this low-efficiency emission in the existence of other broad features, we used low-power excitation and long integration times (60 sec). The background corrected integrated intensities with increasing pump power shows saturation, shown in Figure \ref{fig3}.b The observation of two-electron satellite emission supports the presence of an electron ground state, hence potential spin qubit.

\begin{figure}
	\centering
	\includegraphics[width=\columnwidth]{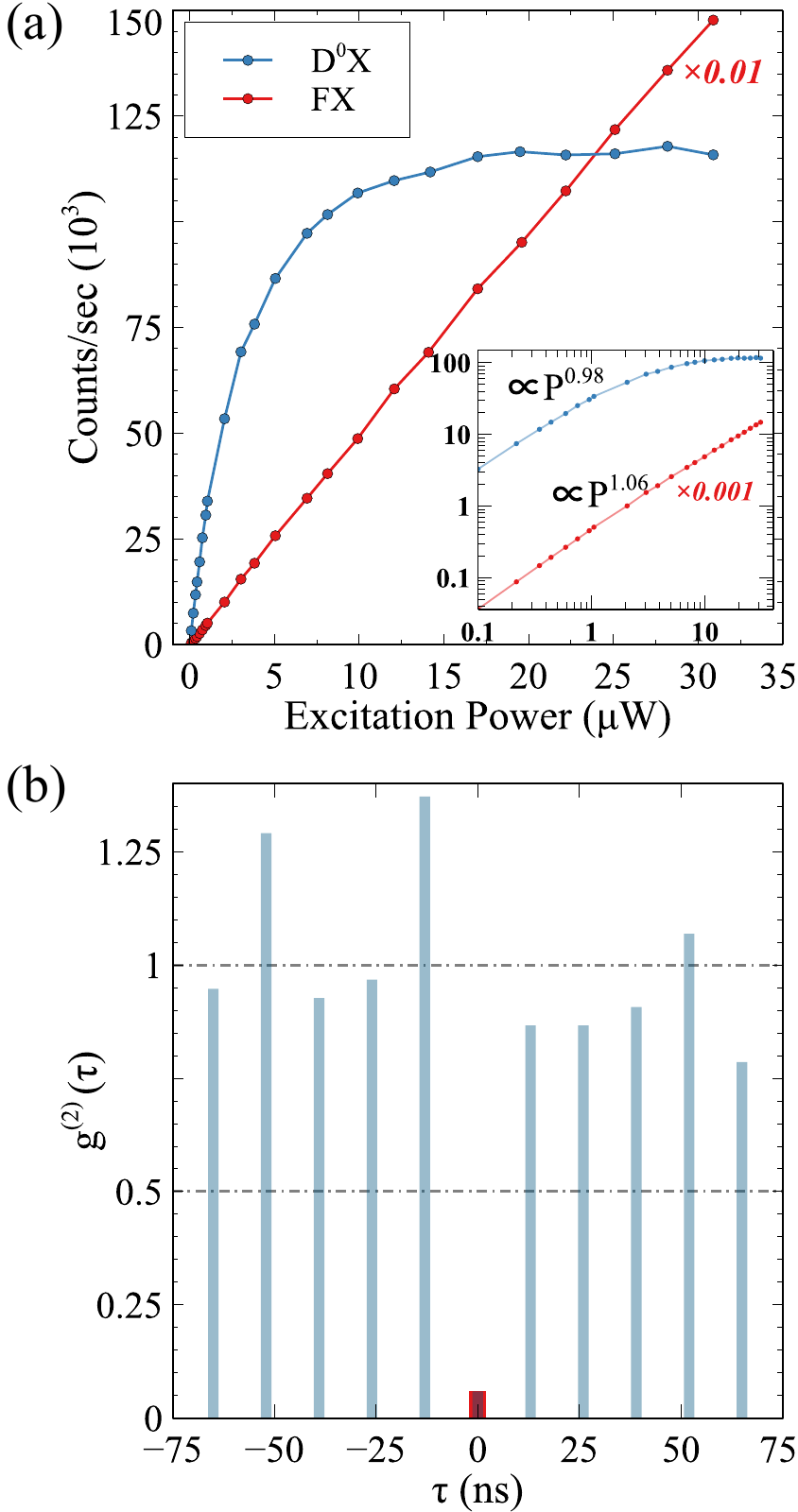}
	\caption{(a) Power dependent intensity of bound exciton (blue) and free exciton (red), inset shows a log scale zoomed-in view at lower excitation powers (b) Second-order auto-correlation function obtained from the single line, demonstrating single photon emission from \DOX.}
	\label{fig4}
\end{figure}

To confirm that the emission originates from a two-level system, we perform intensity-dependent photoluminescence measurements. Figure \ref{fig4}.a shows the emission intensity of both the free exciton and bound exciton line as a function of pump intensity. The bound exciton line exhibits an intensity saturation at pump powers above 5 $\mu$W. In contrast, the free exciton line exhibits no saturation behavior. The inset of the figure shows the log scale power dependence of both curves for the low excitation powers, shown up to 20 $\mu$W. Both \DOX~ and \FX~  lines are fitted for the powers up to 3 $\mu$W, showing nearly linear dependence on the excitation power, with P$^{0.98}$ and  P$^{1.06}$, respectively

In order to demonstrate single-photon emission, we perform second-order correlation measurements using a beam splitter and two single photon detectors. To confirm that the emission is due to a single quantum emitter, we studied the photon statistics of the \DOX~ line. Figure \ref{fig4}.b shows the results of the second-order auto-correlation measurements performed on the single line. We used a frequency-doubled Ti:Sapphire laser emitting at 405 nm with a pulse repetition rate of 76 MHz to excite the emitters. Then, we coupled the collected emission to a single mode fiber to have spatial filtering. The spectral filtering is performed by using narrow slits and a monochromator. Noise contributions from unfiltered background emission and detector dark counts are subtracted \cite{Kim2016}. The second-order correlation exhibits clear anti-bunching with g$^2$=0.06, which confirms that the emission line corresponds to a single photon emitter.

\begin{figure*}
	\centering
	\includegraphics{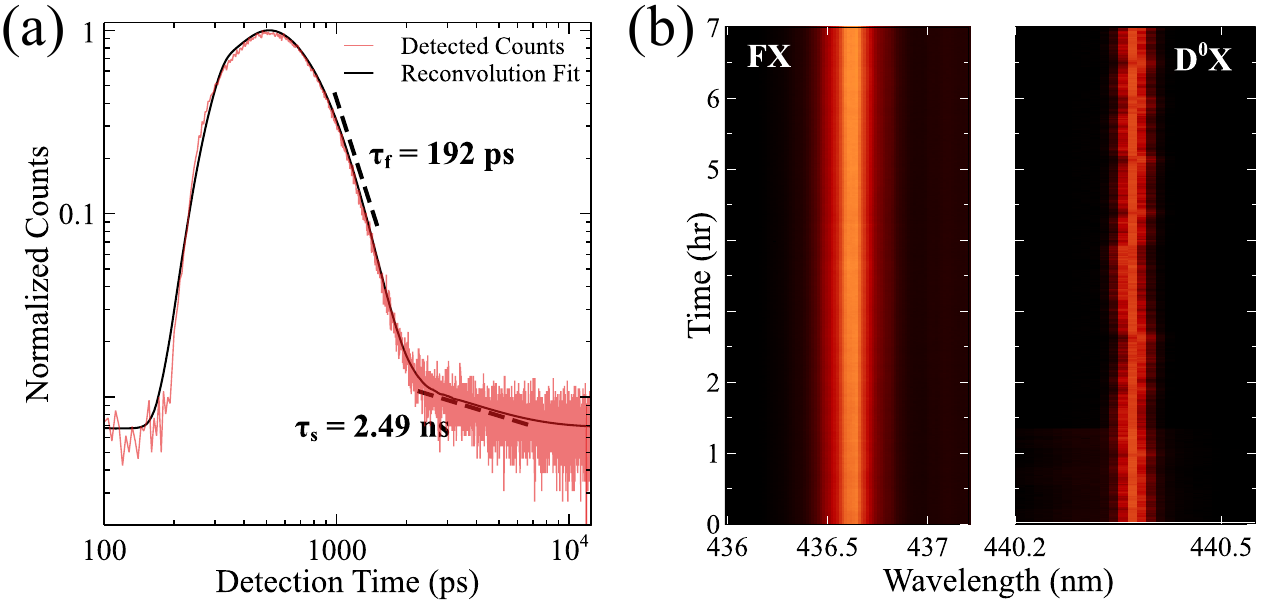}
	\caption{(a) Time-resolved photoluminescence measurement, showing biexponential decay with dominant fast decay component of 192 ps and slow decay component of 2.49 ns, (b) PL intensity recorded over 7 hrs showing the stability of the emitter.}
	\label{fig5}
\end{figure*}

To characterize the donor-bound exciton state's lifetime, we perform time-resolved fluorescence measurements using 3 ps optical pulses. The \DOX~ emission is recorded by a single photon detector with timing jitter on the order of 100 ps. Figure \ref{fig5}.a shows the histogram of photon arrival times obtained from a single bound exciton along with the bi-exponential decay fit. To mitigate the slow timing response of the detectors, we reconvolved (EasyTau software, Picoharp) the detector impulse response recorded at \DOX~ emission wavelength with custom defined decay function.  We extracted the parameters of decay function by fitting the reconvolved function to the measured data. Extracted data shows a biexponential decay, with a dominant fast decay time of 192 ps and a weak contribution from a slower component with a decay time of 2.49 ns. The fast decay time is consistent with previously reported values for other group-7 impurities in ZnSe \cite{Dean1981} and indicates the possibility of realizing a bright and efficient single photon interface with chlorine donor bound excitons in ZnSe. The origin of the slow component may be caused by the repopulation of the \DOX~ state from longer-lived nearby trapping states and is a matter of further investigation.

Figure \ref{fig5}.b highlights the long-term stability of the \DOX~. The figure plots the photoluminescence spectrum over 7 hours under continuous above barrier excitation. The emission does not display any blinking, though slow spectral wandering is observed over a timescale of hours. We calculated the standard deviation of spectral wandering as 0.009 nm, which is in the same order of our spectrometer resolution. Nonexistence of blinking demonstrates of the stability of the emitter.

\begin{figure}
	\centering
	\includegraphics[width=\columnwidth]{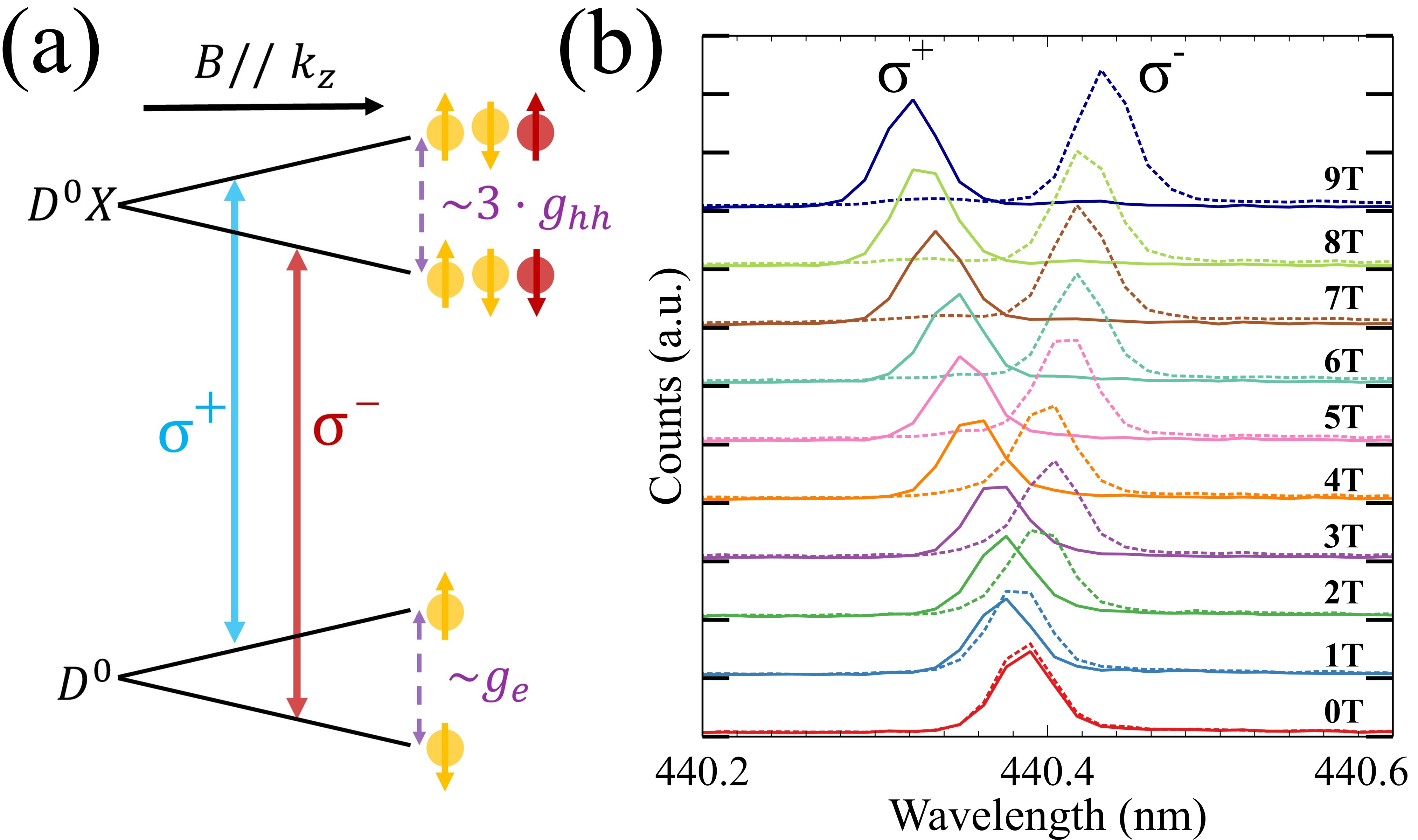}
	\caption{(a) Optical selection rules under applied out of plane magnetic field, (b) Polarization resolved photoluminescence spectrum recorded at increasing magnetic field intensity, showing the splitting of orthogonal circular polarizations.}
	\label{fig6}
\end{figure}

To investigate the spin properties of the bound exciton complex, we performed magneto photoluminescence measurements on the \DOX~ line for magnetic fields up to 9 T applied along the growth direction (Faraday geometry). The optically allowed transitions in this configuration are presented in Figure \ref{fig6}.a. The total spin of the bound exciton ground state is given by the single spin-1/2 electron. In the bound exciton excited state, the donor electron forms a spin-singlet with the electron of the exciton, giving a total spin determined by the spin-3/2 heavy hole \cite{Bayer2002}. For increasing magnetic field, two orthogonally polarized circular emission lines emerge as expected, shown in Figure \ref{fig6}.b. Spectra indicated by solid (dashed) lines are recorded for $\sigma^{+}$ ($\sigma^{-}$) polarization. The energy difference of the two allowed transitions is proportional to $g_{eff}=g_{e}-3g_{hh}$  .  From the observed energy splitting between those two transitions, we infer $g_{eff}=0.9$ which is in good agreement with previous results obtained from single fluorine donors in ZnSe \cite{DeGreve2010}. Additionally, we observe a diamagnetic shift of the \DOX~ emission of 3.25 $\mu$eV/T$^2$. The lack of fine structure splitting and the clear splitting into orthogonal circular polarizations are consistent with the expected Kramer’s degeneracy of the neutral donor state containing a single electron in its ground state \cite{Bayer2002}.

\section{CONCLUSION}

In conclusion, we demonstrated that excitons bound to single Cl donor atoms in ZnSe quantum wells can serve as bright and stable single photon sources. Emission of the bound excitons exhibited clear saturation and single photon emission was confirmed by observing photon antibunching. The fast radiative recombination time below 200 ps demonstrates the possibility of realizing a single photon source with a brightness comparable to state-of-the-art quantum dots. Coupling donor atoms to photonic structures can further decrease the lifetime and lead to high-cooperativity cavities. Observation of two-electron-satellite and circular polarization under Faraday magnetic field indicates a single electron ground state which could be employed for spin-qubit applications. Observation of two-electron-satellite and circular polarization under Faraday magnetic field indicates a single electron ground state which could be employed for spin-qubit applications. Isotopic purification of the host crystal to be nuclear spin free makes Cl impurities an attractive candidate for long-lived memory qubits \cite{Pawlis2019} and providing a route towards optically bright spin qubits with long coherence times. These unique aspects could allow a wide range of applications in quantum information processing.

\begin{acknowledgments}
This work is supported by the Air Force Office of Scientific Research (grant \#FA95502010250), The Maryland-ARL Quantum Partnership (grant \#W911NF1920181) and Deutsche Forschungsgemeinschaft (DFG, German Research Foundation) under Germany’s Excellence Strategy-Cluster of Excellence Matter and Light for Quantum Computing (ML4Q) (EXC 2004 1-390534769). R.M.P acknowledges support through an appointment to the Intelligence Community Postdoctoral Research Fellowship Program at the University of Maryland, administered by Oak Ridge Institute for Science and Education through an interagency agreement between the U.S. Department of Energy and the Office of the Director of National Intelligence.
\end{acknowledgments}

\bibliography{References.bib}

\end{document}